\documentstyle[ltwol,psfig]{article}

\def\Journal#1#2#3#4{{#1} {\bf #2}, #3 (#4)}

\def\NPB{{\em Nucl. Phys.} B}
\def\NPS{{\em Nucl. Phys.} (Proc.Suppl.)}

\def\PRD{{\em Phys. Rev.} D}
\def\RMP{\em Rev. Mod. Phys.}

\newcommand{\be}{\begin{eqnarray}}
\newcommand{\ee}{\end{eqnarray}}
\newcommand{\ba}{\begin{array}}
\newcommand{\ea}{\end{array}}

\newcommand{\Op}{{\cal O}}

\newcommand{\K}{{\cal K}}

\begin{document}

\title{$B^0$-$\bar{B}^0$ mixing from lattice QCD}

\author{N. Yamada and S. Hashimoto\\ for the JLQCD collaboration}

\address{High Energy Accelerator Research Organization~(KEK),
         Tsukuba, 305-0801, Japan\\
         E-mail: nyamada@suchi.kek.jp
         }   


\twocolumn[\maketitle\abstracts{
  We report on lattice calculations of the $B$ meson decay
  constant and the $B$-parameters using the NRQCD action for heavy
  quark. 
  We performed a set of calculations at three lattice spacings in the
  quenched approximation in order to investigate the systematic
  uncertainty associated with the lattice discretization.
  Our results indicate that the systematic error is under good control
  for $f_B$ and $B_B$.
  We also present preliminary results from our unquenched
  simulation. 
}]

\section{Introduction}

The mass difference $\Delta M_q$ in the neutral $B_q$ ($q$=$s$ or $d$)
meson system may be used to determine the CKM elements $|V_{td}|$ or
$|V_{ts}|$, and could eventually be a probe to explore new physics
beyond the Standard Model.
Experimental effort to measure these quantities has already reached 
excellent accuracy for $\Delta M_d$~\cite{dm_exp}, and
$\Delta M_s$ is expected to be observed in the near future.
In order to determine $|V_{tq}|$ and then to search for the new
physics effects using the mass difference, one has to know the matrix
element of the four-quark operator 
\begin{equation}
  \Op_{L_q}
  =
  \bar{b}\gamma_\mu (1-\gamma_5) q\ 
  \bar{b}\gamma_\mu (1-\gamma_5) q,
\end{equation}
with $B_q^0$ and $\bar{B}_q^0$ mesons as external states.
Conventionally, the matrix element is parameterized in terms of the
leptonic decay constant $f_{B_q}$ and the $B$-parameter $B_{B_q}$ as 
\begin{equation}
  \langle \bar{B}_q^0|{\Op}_{L_q}(\mu_b) | B_q^0\rangle
  =
  \frac{8}{3} f_{B_q}^2 B_{B_q}(\mu_b) M_{B_q}^2,
\end{equation}
when the operator $\Op_{L_q}$ is defined at scale $\mu_b$.
The mass difference is, then, given by~\cite{BBL}
\begin{eqnarray}
  \Delta M_{B_q}
  &=&
  \K \cdot|V_{tb}^*V_{tq}|^2
  \left[ \alpha_s(\mu_b) \right]^{-6/23}
  \left[ 1 + \frac{\alpha_s(\mu_b)}{4\pi}J_5 \right]
  \nonumber\\
  & &
  \times\frac{8}{3} f_{B_q}^2 B_{B_q}(\mu_b) M_{B_q}^2,
\label{eq:mass_difference}
\end{eqnarray}
with $\K$ a known factor.
The scale $\mu_b$ is usually set to around $m_b$.

In this report we present recent results from a lattice QCD
calculation of the $B$ meson leptonic decay constant and the 
$B$-parameter obtained by the JLQCD collaboration.
A part of this study has already been published
in~\cite{fB_vpp,bb_paragon_1,bb_paragon_2,bb_vpp}.

It is also possible to estimate the width difference of $B_s$ mesons 
using the heavy quark expansion and lattice QCD.
Our attempt is also reported at this conference~\cite{bcp4_shoji}.

\section{Lattice simulations}
To obtain reliable results from lattice calculations,
systematic uncertainty has to be under control.
To disentangle possible systematic errors, we perform three sets of
lattice simulations, which we call ``Nf0-A'', ``Nf0-B'' 
and ``Nf2'', respectively.
The lattice parameters for these three sets are summarized in
Table~\ref{tab:sim_paras}. 
\begin{table*}
\begin{center}
\begin{tabular}{c|ccc|c|c}
\hline
Run        & \multicolumn{3}{c|}{Nf0-A} & Nf0-B & Nf2 \\
\hline
$N_f$      & \multicolumn{3}{c|}{0} & 0 & 2 \\
$\beta$    & 6.1 & 5.9 & 5.7 & 6.0 & 5.2\\
lattice size
           & 24$^3\times$64 & 16$^3\times$48 & 12$^3\times$32
           & 20$^3\times$48 & 20$^3\times$48 \\
$a$ [fm]   & 0.086 & 0.120 & 0.176 & 0.108 & 0.095 \\
\# of conf.& 518 & 419 & 420 & 180 & 3000 traj.\\
\hline
\end{tabular}
\caption{
  Simulation parameters.
  \label{tab:sim_paras}.}
\end{center}
\end{table*}
In these runs the physical volume of the lattice is kept at about 
(2~fm)$^3$. 
Run ``Nf0-A'' consists of three quenched lattices with different
lattice spacings, which may be used to study scaling behavior
of the decay constant and the $B$-parameter.
Runs ``Nf0-B'' and ``Nf2'', both of which are still preliminary, have 
similar parameters except for the number of dynamical flavors.
We use these runs to investigate the sea quark effect.

In this work we use the lattice NRQCD action~\cite{nrqcd} including
entire $1/m$ corrections for heavy quark.
This approach has an advantage that the $b$ quark can be directly
simulated and then extrapolation in heavy quark mass is not necessary. 
The light quark is described by the $O(a)$-improved Wilson quark 
action~\cite{SW} and, in particular, in Runs ``Nf0-B'' and ``Nf2''
the $O(a)$-improvement is done nonperturbatively.

Since the light quark much lighter than the strange quark can not be 
treated on the lattice with reasonable computer time, the simulations
are carried out at four or five different light quark masses around
and above the strange quark mass, and the results are extrapolated to
the physical light quark mass. 
In particular, in the dynamical quark simulations ``Nf2'', independent 
generations of gauge configurations are necessary to simulate
different sea quark masses.

\section{Decay constant}
\subsection{quenched results}

\begin{figure}
\psfig{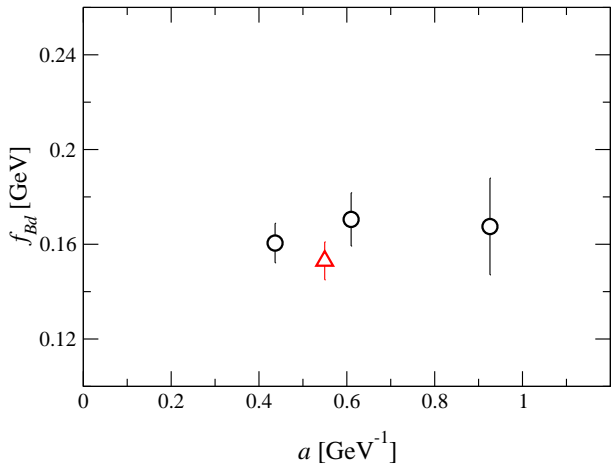}
\caption{
  Lattice spacing dependence of the decay constant in the quenched
  approximation.
  Error bars include statistical and systematic errors except for the
  quenching.
  Open circles are from our previous work.
  Open triangle is a new result with nonperturbatively $O(a)$-improved
  light quarks (Run ``Nf0-B'').
  \label{fig:fb_quench}}
\end{figure}

Several years ago, we performed a lattice calculation
of the $B$ meson decay constant using the NRQCD action~\cite{fB_vpp}. 
The calculations in the quenched approximation were carried out at
three lattice spacings to investigate the scaling property.
The results are plotted in Figure~\ref{fig:fb_quench} (open circles). 

At the present work, in Run ``Nf0-B'' we performed another quenched
calculation, in which the light quark action is nonperturbatively
$O(a)$-improved in contrast to the previous study, in which the
improvement is done at the one-loop level.
The result is also plotted in Figure~\ref{fig:fb_quench} (open
triangle). 
We observe from this figure that within the given uncertainty results
are consistent with each other. 
It indicates that the systematic uncertainty depending on the lattice
spacing is well controlled with the use of the $O(a)$-improved
actions.

There are many other calculations of the $B$ meson decay constant in
the quenched approximation, for which recent summaries are presented
in~\cite{recent_reviews}. 
It is remarkable that results from many groups with different
approaches for heavy and light quarks are in reasonable agreement,
including our results with the NRQCD.

\subsection{unquenching effects}

Last year, the JLQCD collaboration started a new project,
namely QCD simulations with two flavors of dynamical quarks.
A preliminary report on this work is found in~\cite{nf2_jlqcd}.
The calculation of the $B$ meson decay constant is in progress
on these simulations, which we call Run ``Nf2'' in this report. 
Unfortunately, at this stage we have not yet accumulated enough
statistics to draw definite conclusion, so we present a status
report of this work.
All the results presented here are very preliminary.

\begin{figure}
\psfig{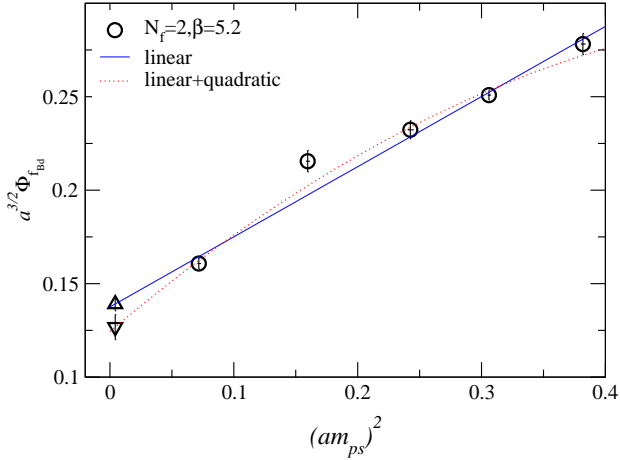}
\caption{
  Chiral behavior of the heavy-light decay constant from Run ``Nf2''. 
  The solid and dotted lines correspond to the linear and quadratic 
  fits, respectively.
  \label{fig:chiral_extrp}}
\end{figure}

Because of the lack of statistics, the chiral limit of
\begin{equation}
  \Phi_{f_{B_d}}
  =\left(\frac{\alpha_s(M_{P_d})}{\alpha_s(M_{B_d})}
                 \right)^{2/\beta_0}f_{P_d}\sqrt{M_{P_d}}
\end{equation}
still has large uncertainty as shown in
Figure~\ref{fig:chiral_extrp}. 
There we attempt various fit functions to obtain the physical light
quark mass limit, and linear (solid) and quadratic (dotted) fits, are
shown in Figure~\ref{fig:chiral_extrp}. 
Both of them have large $\chi^2/dof$ and the chiral limit is not very
trusted. 
Perhaps, the reason for the bad statistical behavior of the unquenched 
calculation is that the simulation has long autocorrelation in the
direction of fictitious time in the molecular dynamics evolution.
More work is necessary to understand it and eventually to perform
reliable chiral extrapolation.

\begin{figure}
\psfig{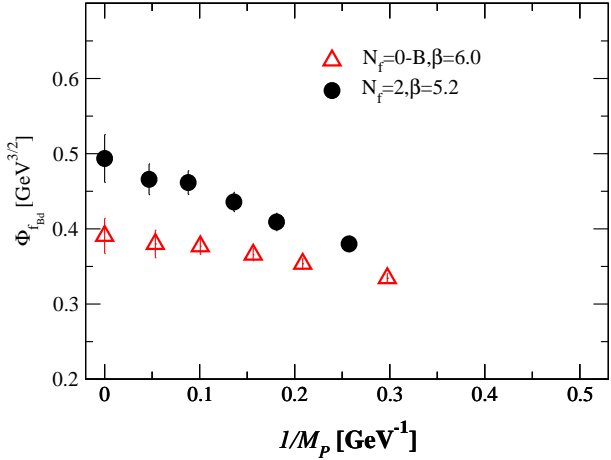}
\caption{
  Comparison of the decay constant from quenched (triangles) and
  unquenched (circles) simulations.
  \label{fig:fb_q_unq}}
\end{figure}

In spite of the poor data, especially toward the chiral limit, we
attempt to see the effect of unquenching by comparing the quenched and
unquenched calculations of the decay constant.
In Figure~\ref{fig:fb_q_unq}, the unquenched result (circles) from Run 
``Nf2'' are compared with the quenched one (triangles) from Run
``Nf0-B''. 
For the unquenched data, we use the result of quadratic fit in the
chiral extrapolation.
It seems evident that the unquenched result is higher than the
quenched one, as recently found by several
groups~\cite{recent_reviews}.
More work (and statistics) is, however, necessary to draw definite
conclusion. 
Results with higher statistics will appear in the summer conferences
in this year.

\section{$B$-parameter}

\subsection{quenched results}

\begin{figure}
\psfig{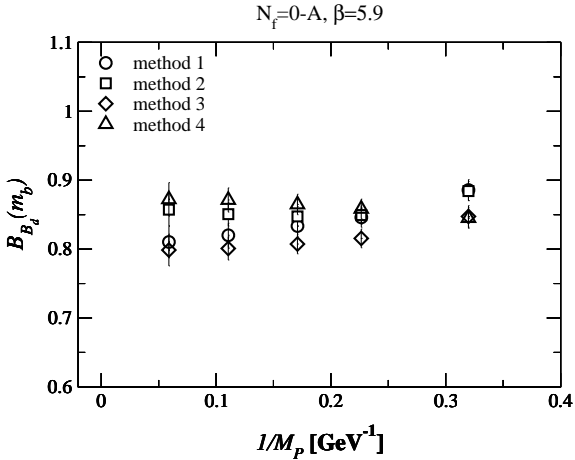}
\caption{
  $B_{B_d}(m_b)$ as a function of $1/M_P$ at $\beta$=5.9 
  in Run ``Nf0-A''.
  Four symbols correspond to four different methods adapted.
  \label{fig:mass_dep_BB_59}}
\end{figure}

\begin{figure}
\psfig{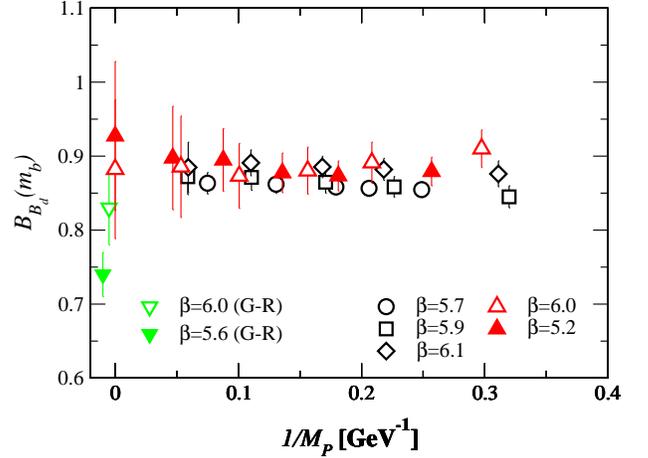}
\caption{
  $B_{B_d}(m_b)$ as a function of $1/M_P$ at three lattice spacings
  (from Run ``Nf0-A'') .
  The method 4 is used.
  The open and filled triangles in the static limit come
  from the lattice HQET calculations by Gim\'enez and Reyes with
  $N_f$=0 and 2, respectively \protect\cite{Gimenez_Reyes_01}.
  \label{fig:mass_dep_BB_all}}
\end{figure}

Quenched results for $B_B$ on the Run ``Nf0-A'' were already reported
in \cite{bb_vpp}. 
Figure~\ref{fig:mass_dep_BB_59} shows the $1/M$ dependence of
$B_{B_d}(m_b)$ obtained at $\beta$=5.9.
Four symbols represent four different methods to calculate the
$B$-parameters as discussed in \cite{bb_vpp}, and those have different
systematic errors at order $\alpha_s\Lambda_{\rm QCD}/m$ or at order
$\alpha_s^2$.
We may use the difference among the methods to estimate the size of
systematic errors.

In Figure~\ref{fig:mass_dep_BB_all}, we plot all the results from Run
``Nf0-A'', where open circles, boxes and diamonds represent
the results at $\beta$=5.7, 5.9 and 6.1, respectively.

We find that the results from different lattice spacings agree
perfectly, suggesting that the systematic error associated with the
lattice discretization is small.
Another result from Run ``Nf0-B'' is also plotted in the same figure 
(open triangle-up), and we find a nice agreement again.

The open triangle-down at $1/M_P$=0 is a result from the static
approximation by Gim\'enez and Reyes~\cite{Gimenez_Reyes_01},
which is consistent with our result extrapolated to the static limit. 
\footnote{
  We also note that the APE~\cite{bb_ape} and UKQCD~\cite{bb_ukqcd} 
  collaboration have recently calculated the $B$-parameters using the 
  relativistic lattice action for heavy quark, and the results are
  consistent with ours.
  Their approach has completely different systematic errors
  growing as a power of $aM$, heavy quark mass in the lattice unit. 
  Therefore, an extrapolation is necessary from lighter heavy quark
  mass region to $m_b$.
  }

\subsection{unquenching effects}
Using the same technique we developed in the quenched approximation,
we are calculating the $B$-parameters including the effect of
dynamical quarks in the Run ``Nf2''.
The problem of ill-behaved chiral extrapolation does not exist for
$B_B$, since the light quark mass dependence of the $B$-parameters
is very small.
Our preliminary result for $B_{B_d}(m_b)$ is shown in
Figure~\ref{fig:mass_dep_BB_all} (filled triangle-up), where we do not
find significant effect of the unquenching.

A previous unquenched calculation by Gim\'enez and Reyes
\cite{Gimenez_Reyes_01} is also plotted in the figure (filled
triangle-down). 
There result seems significantly lower than others, but the
calculation is done without the $O(a)$-improvement and the result
should have different systematic uncertainty.

Our preliminary results for $B_B$ are 
\begin{eqnarray}
    B_{B_d}(m_b)&=& 0.83(3)(8), \\
 B_{B_s}/B_{B_d}&=& 1.01(2)(2).
\end{eqnarray}

\section{Applications}
Let us briefly discuss the application of our results to the
phenomenology.

Substituting our results for $B_{B_d}(m_b)$ and the current world average
for $f_{B_d}$ by C. Bernard~\cite{recent_reviews}
to the formula of mass difference Eq.~(\ref{eq:mass_difference}),
we obtain
\begin{equation}
0.71< \left|\frac{V^*_{tb}V_{td}}{V^*_{cb}V_{cd}}\right| <1.01.
\end{equation}

Although the mass difference of $B_s$ mesons has not been discovered
yet, precise measurement is expected in the near future at Tevatron Run II.
We predict it as
\begin{equation}
  \Delta M_s = 18.2(4.7)~{\rm ps}^{-1}.
\end{equation}
The current experimental lower bound~\cite{dm_exp} is $15.0$ ps$^{-1}$.
In the previous report~\cite{bb_vpp}, we quoted higher value 
$\Delta M_s=20.0(4.7)$ {\rm ps}$^{-1}$.
That is because the world average for $f_{B_s}$ changed since that time.

\section{Conclusion}
It is essential for the determination of $|V_{td}|$ to calculate the 
$B$ meson decay constant and the $B$-parameters with controlled
errors.
The systematic errors associated with the lattice discretization are
proved to be under good control in our calculation using the NRQCD
action for heavy quark.
Both for $f_B$ and $B_B$ we show that the scaling violation is small
in the quenched approximation.

Moreover, we started a simulation including the effect of dynamical
quarks. 
At the moment, our statistics is not enough to draw definite
conclusion, but we expect results in the near future.

\section*{Acknowledgments}

We thank Ken-Ichi Ishikawa, Takashi Kaneko, Tetsuya Onogi and
other members of JLQCD collaboration for useful discussions.
This work is supported by the Supercomputer Project No. 54 (FY2000)
of High Energy Accelerator Research Organization (KEK), also
in part by the Grant-in-Aid of the Ministry of Education (No. 11740162).
N.Y. is supported by the JSPS Research Fellowship.

\end{document}